\documentclass[12pt]{article}
\usepackage{graphicx}
\usepackage{epsfig}
\usepackage{amsbsy}

\begin{document}
\baselineskip 10mm

\centerline{\large \bf On the Temperature Dependence of the Lifetime}
\centerline{\large \bf of Thermally Isolated Metastable Clusters }

\vskip 6mm

\centerline{L. A. Openov$^{*}$, D. A. Lobanov, and A. I. Podlivaev}

\vskip 4mm

\centerline{\it Moscow Engineering Physics Institute (State
University), 115409 Moscow, Russia}

\vskip 2mm

$^{*}$ E-mail: LAOpenov@mephi.ru

\vskip 8mm

\centerline{\bf ABSTRACT}

The temperature dependence of the lifetime of the thermally isolated metastable N$_8$ cubane up to
its decay into N$_2$ molecules has been calculated by the molecular dynamics method. It has been
demonstrated that this dependence significantly deviates from the Arrhenius law. The applicability
of the finite heat bath theory to the description of thermally isolated atomic clusters has been
proved using statistical analysis of the results obtained.

\vskip 15mm


\newpage

Simulations of the time evolution of atomic clusters at high temperatures are of interest from
several points of view. An analysis of distortions of the cluster structure and products of its
decay can provide information on the mechanism of the final stage of the cluster formation and
possible components for its synthesis, respectively. For example, when the C$_{60}$ fullerene is
heated, defects are formed on its {}``surface'' because of the Stone-Wales transformations, while
the formation of the fullerene from an initial imperfect C$_{60}$ cluster occurs due to the same
transformations going in the reverse order [1, 2]. In turn, the decomposition of the hydrocarbon
cubane C$_8$H$_8$ into benzene and acetylene molecules suggests the use of these compounds for the
synthesis of the cubane (applying catalysts to speed up the reverse reaction) [3].

The main parameter determining the thermal stability of a metastable cluster is the activation
energy of its decomposition (or the transformation into another isomer) $E_a$. If the height of the
minimum energy barrier preventing the decomposition is $U$, the inequality $E_a > U$ holds true. If
there is only one decomposition channel, the equality $E_a = U$ is satisfied. The magnitude of $E_a$
is determined [4] in experiments by measuring the lifetime $\tau$ of the cluster at different
temperatures $T$ and making use of the Arrhenius law
\begin{equation}\label{1}
 \tau^{-1}(T)=A\exp(-E_a/k_BT), 
\end{equation}
where $A$ is the frequency factor and $k_B$ is the Boltzmann constant. The theoretical dependence of
$\tau$ on $T$, in principle, can be determined by numerical simulations of the cluster lifetime
using the molecular dynamics method (the so-called {}``computer experiment''). 

However, it should be noted that Eq. (1) is valid only when the cluster is in thermal equilibrium
with the environment (it is thermalized). In the experiment, the thermal contact is ensured by
placing clusters in a {}``buffer gas'' (usually, helium is used) heated to a required temperature
[4]. In the theory, the canonical $NVT$ ensemble is used to describe thermalized clusters. However,
sometimes cluster disassociates so quickly that it does not have enough time to come into thermal
equilibrium with its environment. It happens, for example, during photofragmentation, when the
clusters excited by a laser pulse do not collide with each other for a time $t < 1 \mu$s required
for their decomposition. In this case, the dynamics of the cluster is simulated using the
microcanonical $NVE$ ensemble [5] and the total energy of the cluster $E=E_{kin}+E_{pot}$ (the sum
of the kinetic and potential energies) remains constant during its evolution. Here, the role of the
temperature is played by the so called {}``microcanonical'' (or {}``dynamic'') temperature $T_m$,
which is a measure of the kinetic energy of relative motion of atoms in the cluster, which is at
rest and does not rotate as a whole (i.e., it is basically the measure of the excitation energy of
the cluster). It is calculated using the formula [6, 7]
$\langle E_{kin}\rangle=\frac{1}{2}k_BT_m(3n-6)$, where $\langle E_{kin}\rangle$ is the
kinetic energy of the cluster averaged over the microcanonical ensemble (or over the time) and $n$
is the number of atoms in it.

The question arises: is it possible to use Eq. (1) to determine the decomposition activation energy
of a thermally isolated cluster by simply substituting $T$ by $T_m$ in it? According to the so-called
finite heat bath theory [8, 9], in which the cluster itself plays the role of the reservoir for the
degree of freedom along the reaction coordinate that leads to the decomposition, this is
possible. But there is a reservation: the equation should include the quantity
$T^*=T_m-E_a/2C $ instead of $T_m$, where $C=(3n-6)k_B$ is the specific heat of the cluster.

In order to experimentally test this assertion, it is necessary, first, to ensure that there is
only one channel of decomposition (otherwise, the dependences of $\tau$ on $T$ and, correspondingly,
$\tau$ on  $T_m$ become more complex) and, second, to know $E_a$ for the decomposition in the
channel. However, usually the cluster has several decomposition channels (sometimes they are
completely different [10]), and even if the decomposition goes mainly over one of them, the
corresponding activation energy is not known a priori. Since, for medium-sized clusters and
temperatures, at which the cluster decomposition time is $t < 1 \mu$s, the correction $E_a/2CT_m$
is usually very small ($\approx 0.07$ for C$_{60}$ fullerene [2]), the use of $T^*$ instead of $T_m$
in Eq. (1) does not lead to any substantial change in the function $\tau (T_m)$, even though it
leads to somewhat better agreement of numerical simulations with experiment [2].  

The purpose of the present paper is to directly verify the finite heat bath theory via numerical
simulations of dynamics of a metastable cluster for which there is only one decomposition channel
and the height of the corresponding potential barrier is known from independent calculations. We
choose the N$_8$ cubane (Fig. 1), which has been predicted theoretically [11-13] but, as far as we
know, has not been observed in experiment yet, to play the role of such a cluster. Our choice is
stipulated mainly by a small number of atoms in the cubane (and, correspondingly, a comparatively
large magnitude of the correction to microcanonical temperature), and also by the fact that the
products of its decomposition (four N$_2$ molecules) are known, which simplifies the search for
barriers that prevent the decomposition (the more so because the initial cluster has a high
symmetry). Another reason for the choice of so exotic cluster for testing of the statistical laws
is irreversible character of decomposition of the cubane after going over the activation barrier
nearest to the local energy minimum. The majority of clusters do not possess this feature and come
from the initial configuration to adjacent metastable configurations and back many times during
their evolution, because of which it is quite difficult to determine the lifetime of the cluster
[2, 4].

We calculated the interatomic distance $a$ (the length of the N-N bond) and the disassociation
energy of the cubane $E_{diss}=E($N$_8)-4E($N$_2$) using the density functional theory with the
6-31G$^*$ basis set and the exchange-correlation functional B3LYP. Our results
($a=1.52$ \AA, $E_{diss}=2.27$ eV/atom) agree with the data of other authors [11-13] obtained using
different ab initio methods ($a=1.46-1.54$ \AA, $E_{diss}=1.4-3.1$ eV/atom). We also found the
reaction path for the cubane decomposition along the reaction coordinate, which connects the
initial (N$_8$ cubane) and final (four N$_2$ molecules) configurations in the space of atomic
coordinates. It turned out that the path contains several barriers that correspond to sequential
transitions of the N$_8$ cluster to different intermediate configurations (Fig. 2). Each barrier
has the corresponding saddle point at the potential energy hypersurface. Maximum energy corresponds
to the saddle point of the first (the nearest to the cubane) barrier. Therefore, this barrier
determines the stability of the cubane. Its height is $U=0.88$ eV. We obtained similar results
using the Hartree-Fock method and the Moller-Plesset second-order perturbation theory (in
particular, $U=0.80$ eV).

In principle, at this stage, we can proceed to the determination of the dependence of the lifetime
of the cubane $\tau$ on the microcanonical temperature $T_m$ by setting different initial
excitation energies for the cluster and employing the molecular dynamics method to analyze its
evolution. However, the ab initio calculations require very large amounts of the computer time,
because of which the {}``life'' of the cluster can be observed only over a very short period of time
(1 - 10 ps), which is insufficient to collect necessary statistics. Therefore, we chose another way:
we found the tight binding potential for nitrogen systems. It is known that tight binding models
produce a reasonable compromise between rigorous ab initio approaches and oversimplified classic
potentials of the interaction between atoms. Often, they incur only slight loss in accuracy as
compared to the former approaches (because they take into account explicitly the quantum-mechanical
{}``band'' contribution of the electron subsystem to the total energy) [1, 14] and simultaneously
make it possible to perform the {}``computer experiment'' for macroscopic (on atomic scales) times
$\sim 1 \mu$s [3, 15].

We start from the nonorthogonal tight binding potential [16], which we successfully used before
to simulate dynamics of the hydrocarbon cubane C$_8$H$_8$ [3, 15]. We choose parameters of the
potential for the case of nitrogen systems to achieve the best possible correspondence of
interatomic distances, binding energies, and minimum oscillation frequencies to experimental values
(for N$_2$ molecules) or to results of ab initio calculations (for the N$_8$ cubane and some other
metastable nitrogen clusters). In particular, for the case of the cubane, our model leads to
$a=1.54$ \AA, and $E_{diss}=2.22$ eV/atom. The analysis shows that the dependence of the potential
energy of the N$_8$ system on the reaction coordinate along the decomposition path of the cubane
has the same form as in Fig. 2; i.e., in agreement with the ab initio calculations, the stability
of the cubane is determined by the height of the nearest barrier. It turns out to be 0.384 eV within
the tight binding model, which is approximately two times smaller than the result of our density
functional theory calculations. It is not surprising because it is known that the ab initio
methods, in general, and the density functional theory, in particular, tend to overstimate
heights of potential barriers [3]. However, at present, we are interested not in which of the
theoretical approaches provides more precise calculations, but in the fact that, now, using the
tight binding potential and the value of $U$ determined using the same potential, we can study the
dynamics of the cubane to its decomposition over a very wide range of durations $10^{-13}-10^{-6}$ s
and then compare the temperature dependences of the lifetime of the cubane both with the usual
Arrhenius law and with predictions of the finite heat bath theory [8, 9]. The criterion of the
validity of one or another approach is coincidence of the activation energy found from the
{}``computer experiment'' with the barrier height $U$. 

We simulate the dynamics of the thermally isolated N$_8$ cubane in the following way. At the
initial instant of time, we set each atoms in motion in such a way that the total momentum and the
angular momentum of the cluster as a whole are equal to zero (a special case of the $NVEPJ$
ensemble [15]). Then, forces that act on atoms are calculated. The classical Newton equations
are solved numerically using the velocity Verlet algorithm. The time step used is
$t_0=2.72\cdot 10^{-16}$ s. 

We study the evolution of the cubane for 74 different sets of initial velocities and displacements
of atoms that correspond to the initial microcanonical temperatures $T_m=350-1000$ K. According to
the ergodic hypothesis, $T_m$ is equal to the {}``dynamic'' temperature [15], which is determined
by averaging the kinetic energy of the cluster over 1000-10000 steps of the molecular dynamics
simulation [6, 7]. We found that the cubane decomposes into either four N$_2$ molecules or one N$_2$
molecule and two quasi-linear N$_3$ radicals, each of which, in turn, decomposes into the N$_2$
molecule and the nitrogen atom (according to our calculations, E(N$_3$) $\approx$ E(N$_2$) + E(N), which is
in agreement with results of other authors [17]). The decomposition goes on very quickly
(over 0.1 - 1.0 ps). Since the total energy of the thermally isolated cluster is
$E_{pot}+E_{kin}=$ const, a drastic decrease in the potential energy with the decomposition leads
to the corresponding increase in the kinetic energy, i.e., to the increase in $T_m$ (by several
thousands of degrees). The detailed analysis of the atomic configuration at the initial stage of
the decomposition process shows that it corresponds (with very rare exceptions) to the configuration
of the saddle point in the potential energy of the cluster with the energy barrier $U=0.384$ eV.
When this barrier is overcome, the decomposition becomes irreversible, even though, then, it can go
via different paths (Fig. 2 shows the scheme of the cubane decay into four N$_2$ molecules, but the
final products can also be three N$_2$ molecules and two nitrogen atoms). 

Figure 3 shows the calculated values of ln($\tau$) for different microcanonical temperatures $T_m$.
It can be seen that, over the entire range of $T_m$ under study, the dependence ln($\tau$) on
1/$T_m$ can be fit well by a straight line, which is in agreement with the Arrhenius formula
following for the microcanonical ensemble from Eq. (1) at $T=T_m$. The temperature dependence of
the frequency factor $A$, if any, is so weak that it cannot be revealed using our data. Therefore,
at first glance, the Arrhenius formula describes well the results of the simulations of the
decomposition of the thermally isolated cubane. However, the activation energy determined from the
slope of the approximating straight line in Fig. 3 turns out to be $E_a=0.626\pm 0.022$ eV, which
is much larger than the height $U=0.384$ eV of the barrier that is overcome during the
decomposition. Therefore, the inference can be made that the usual Arrhenius law is inapplicable
to the description of the decomposition of the thermally isolated cluster. 

Let us see what happens with corrections for finite size of heat bath [8, 9]. By assuming that $T$
in Eq. (1) is equal to $T^*=T_m-E_a/2C$ and considering the same set of ln($\tau$) values as a
function of 1/$T^*$ (Fig. 4) instead of the function of 1/$T_m$, we see that ln($\tau$) as a
function of 1/$T^*$ also fits straight line very well, but, in this case,
the activation energy (which was determined by successive iterations) is $E_a=0.367 \pm 0.013$ eV
in excellent agreement with $U=0.384$ eV. Notice that the parameter $E_a/2CT_m$ that describes
magnitude of the correction to $T_m$ is 0.10-0.35 for the range of $T_m$ under study; i.e., it is
substantially larger than that, for example, in simulations of the C$_{60}$ fullerene [2].
Ultimately, this is the fact that allows us to find the clear distinction (much larger than
statistical uncertainty) between the functional dependence of $\tau$ on $T_m$ with and without
corrections for the finite size of the heat bath. 

The frequency factors determined by crossings of the straight lines in Figs. 3 and 4 with the
vertical axis are $A=8.5\cdot 10^{15}$c$^{-1}$ and $4.1\cdot 10^{14}$c$^{-1}$, respectively. It is
worth noting that one and the same set of data can be equally well fit using absolutely different
pairs of fitting parameters ($E_a$, $A$). The sensitivity of the parameters to the form of the
functional dependence $\tau (T_m)$ indicates once again that it is necessary to choose the correct
one to analyze the data on the lifetime of metastable clusters. In the present paper, we directly
prove that the finite heat bath theory is applicable to the quantitative description of thermally
isolated clusters and that it can be used in studies of fast processes of decomposition of clusters
and their interaction with each other.

\newpage
\vskip 20mm
\includegraphics[width=\hsize,height=15cm]{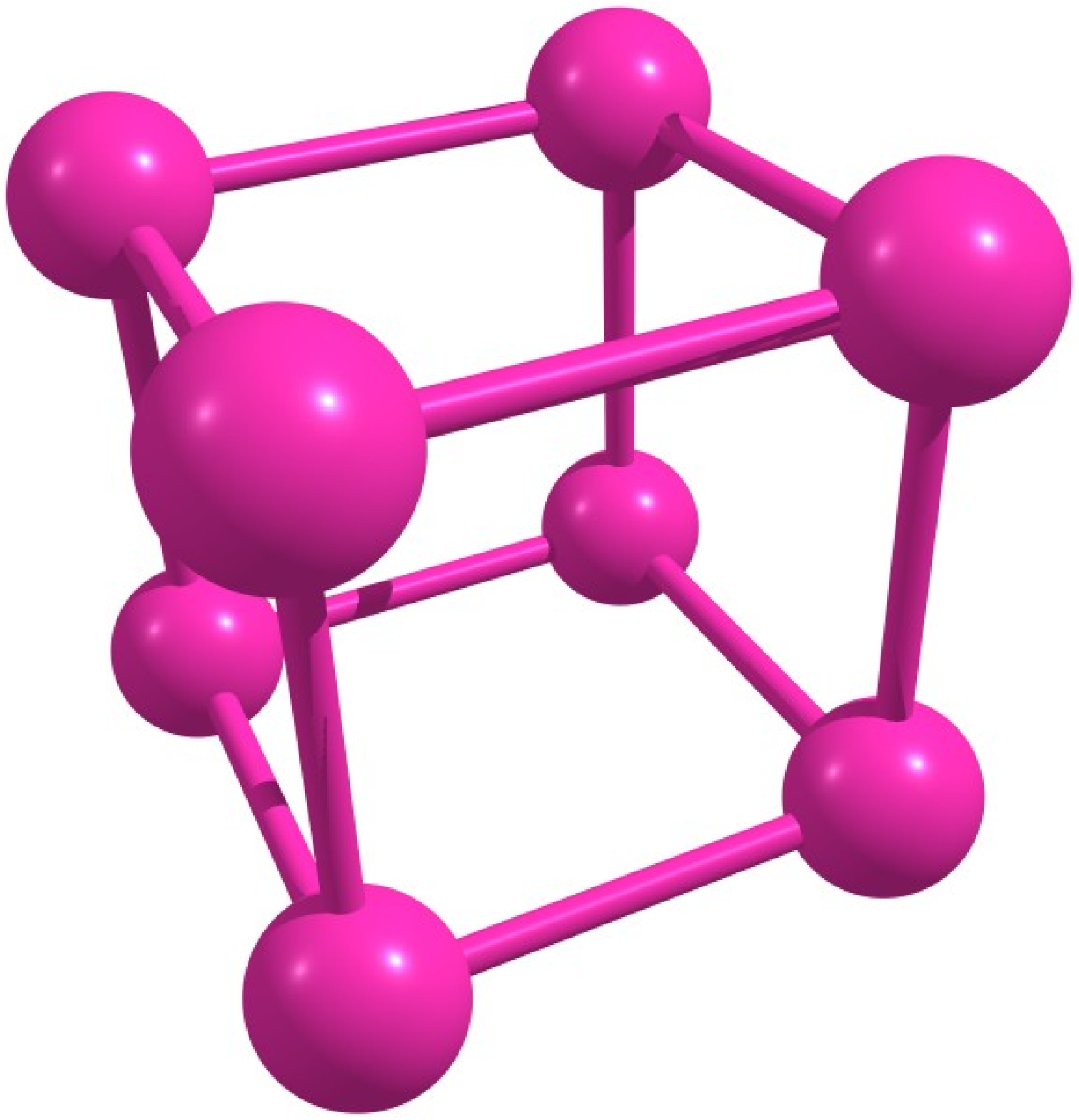}
\vskip 40mm
Fig. 1. N$_8$ cubane.

\newpage
\vskip 20mm
\includegraphics[width=\hsize,height=15cm]{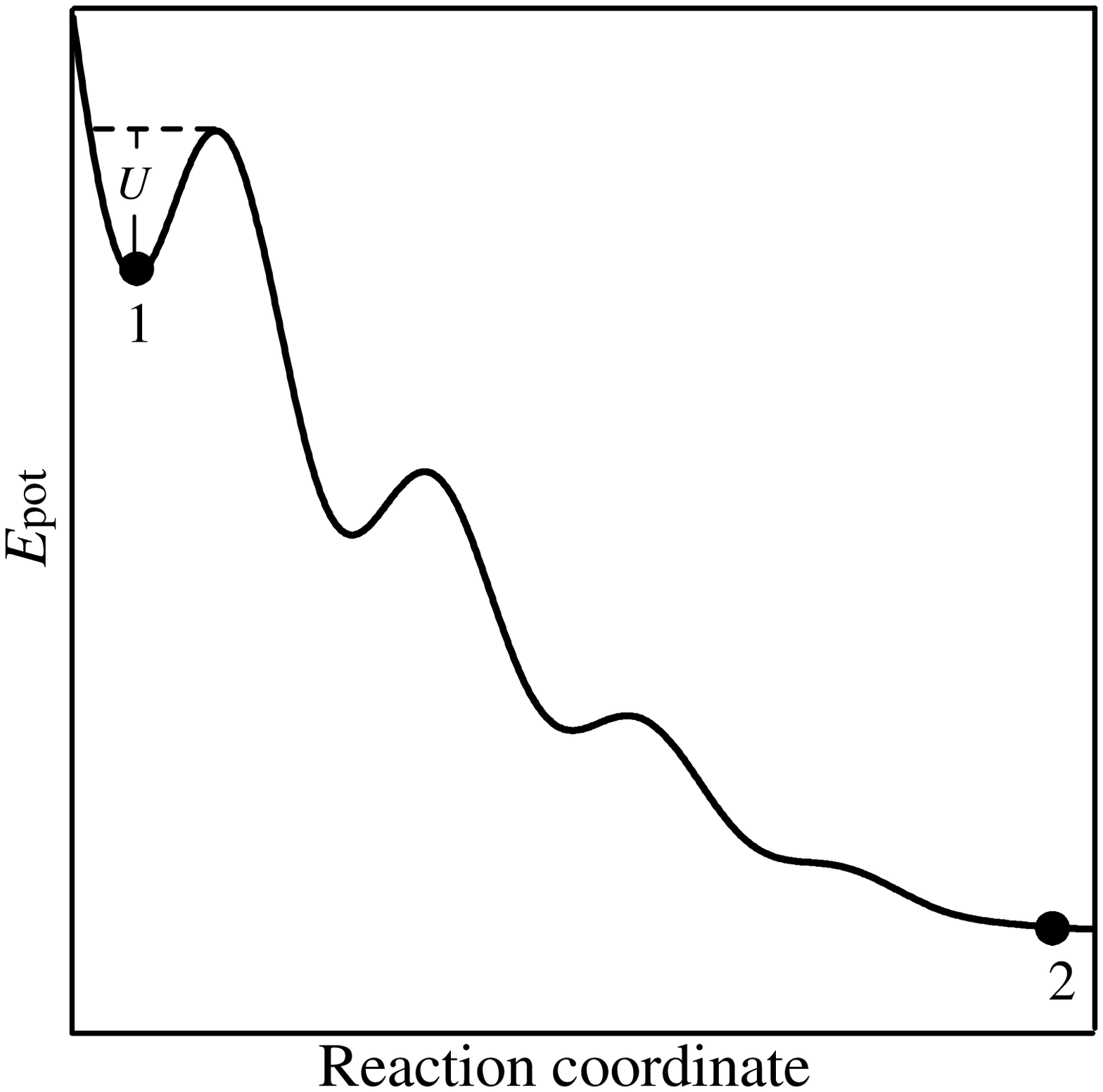}
\vskip 40mm
Fig. 2. Schematic dependence of the potential energy $E_{pot}$ of the N$_8$ cubane (1) on the
reaction coordinate along the path cubane $\rightarrow$ four N$_2$ molecules (2).

\newpage
\vskip 20mm
\includegraphics[width=\hsize,height=15cm]{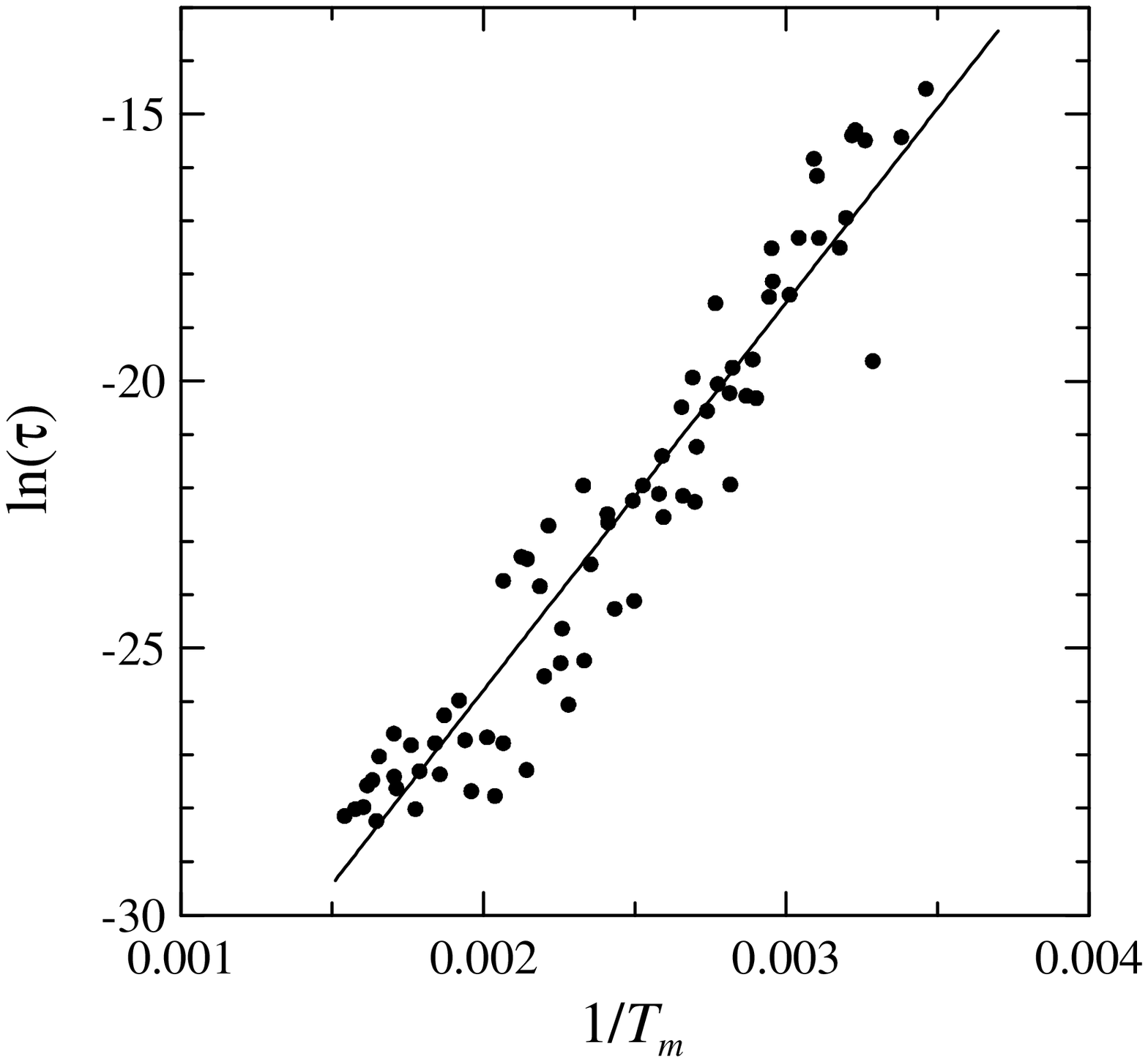}
\vskip 40mm
Fig. 3. Logarithm of the lifetime $\tau$ (in seconds) of the N$_8$ cubane as a function of the inverse
microcanonical temperature 1/$T_m$ (in K$^{-1}$. Points indicate the results of the calculation. The solid line
represents the linear approximation by the least-squares method. 

\newpage
\vskip 20mm
\includegraphics[width=\hsize,height=15cm]{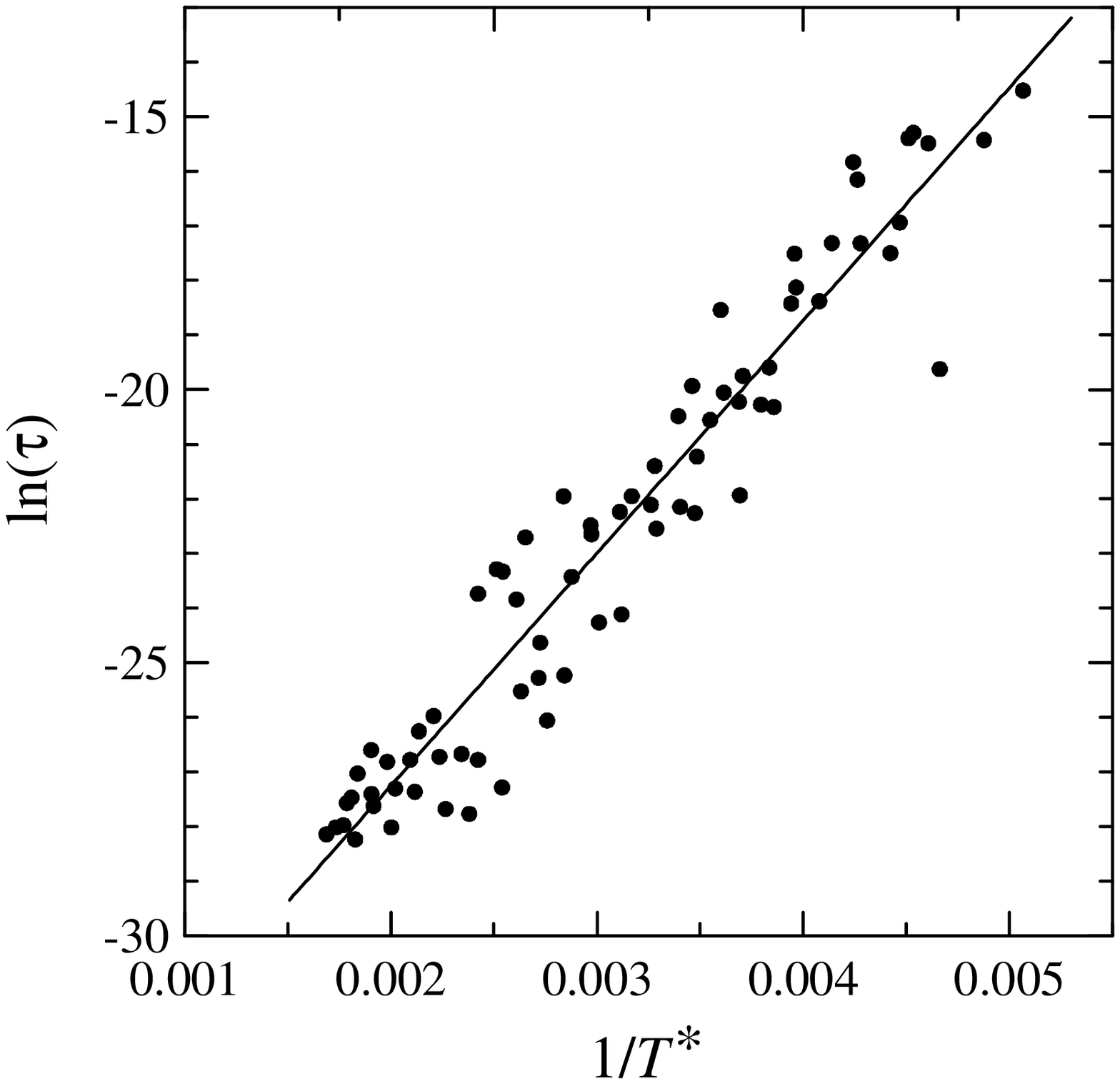}
\vskip 40mm
Fig. 4. The same as in Fig. 3 but in the other coordinates. The quantity inverse to
$T^*=T_m-E_a/2C$ is plotted along the abscissa axis (see text).


\begin{thebibliography}{10}
\bibitem{1} A. I. Podlivaev and L. A. Openov, Pis'ma Zh. Eksp. Teor. Fiz. {\bf 81}, 656 (2005)
[JETP Lett. {\bf 81}, 533 (2005)]. 
\bibitem{2} L. A. Openov and A. I. Podlivaev, Pis'ma Zh. Eksp. Teor. Fiz. {\bf 84}, 73 (2006)
[JETP Lett. {\bf 84}, 68 (2006)]. 
\bibitem{3} M. M. Maslov, D. A. Lobanov, A. I. Podlivaev, and L. A. Openov, Fiz. Tverd. Tela
(St. Petersburg) {\bf 51}, 609 (2009) [Phys. Solid State {\bf 51}, 645 (2009)]. 
\bibitem{4} C.Lifshitz, Int. J. Mass. Spectrom. {\bf 198}, 1 (2000).
\bibitem{5} E. M. Pearson, T. Halicioglu, and W. A. Tiller, Phys. Rev. A {\bf 32}, 3030 (1985).
\bibitem{6} C. Xu and G. E. Scuseria, Phys. Rev. Lett. {\bf 72}, 669 (1994).
\bibitem{7} J. Jellinek and A. Goldberg, J. Chem. Phys. {\bf 113}, 2570 (2000).
\bibitem{8} C. E. Klots, Z. Phys. D {\bf 20}, 105 (1991).
\bibitem{9} J. V. Andersen, E. Bonderup, and K. Hansen, J. Chem. Phys. {\bf 114}, 6518 (2001).
\bibitem{10} L. A. Openov and A. I. Podlivaev, Pis'ma Zh. Eksp. Teor. Fiz. {\bf 84}, 217 (2006)
[JETP Lett. 84 (4), 190 (2006)]. 
\bibitem{11} R. Engelke and J. R. Stine, J. Phys. Chem. {\bf 94}, 5689 (1990).
\bibitem{12} W. J. Lauderdale, J. F. Stanton, and R. J. Bartlett,
J. Phys. Chem. {\bf 96}, 1173 (1992).
\bibitem{13} M. L. Leininger, C. D. Sherrill, and H. F. Schaefer,
J. Phys. Chem. {\bf 99}, 2324 (1995).
\bibitem{14} I. V. Davydov, A. I. Podlivaev, and L. A. Openov, Fiz. Tverd. Tela (St. Petersburg)
{\bf 47}, 751 (2005) [Phys. Solid State {\bf 47}, 778 (2005)]. 
\bibitem{15} L. A. Openov and A. I. Podlivaev, Fiz. Tverd. Tela (St. Petersburg) {\bf 50}, 1146
(2008) [Phys. Solid State {\bf 50}, 1195 (2008)].
\bibitem{16} M. M. Maslov, A. I. Podlivaev, and L. A. Openov, Phys. Lett. A {\bf 373}, 1653 (2009).
\bibitem{17} P. Zang, K. Morokuma, and A. M. Wodtke, J. Chem. Phys. {\bf 122}, 014106 (2005).
\end{thebibliography}
\end{document}